\documentclass[referee]{aa}

\usepackage{psfig}
\begin{document}

\title{BL Lacertae: complex spectral variability and rapid synchrotron flare
detected with $Beppo$SAX}

\author{M. Ravasio\inst{1}
\and G. Tagliaferri\inst{1}
\and G. Ghisellini\inst{1}
\and P. Giommi\inst{2}
\and R. Nesci\inst{3}
\and E. Massaro\inst{3}
\and L. Chiappetti\inst{4}
\and A. Celotti\inst{5}
\and L. Costamante\inst{1}
\and L. Maraschi\inst{6}
\and F. Tavecchio\inst{6}
\and G. Tosti\inst{7}
\and A. Treves\inst{8}
\and A. Wolter\inst{6}
\and T. Balonek\inst{9}
\and M. Carini\inst{10}
\and T. Kato\inst{11}
\and O. Kurtanidze\inst{12}
\and F. Montagni\inst{3}
\and M. Nikolashvili\inst{12}
\and J. Noble\inst{13}
\and G. Nucciarelli\inst{7}
\and C.M. Raiteri\inst{14}
\and S. Sclavi\inst{3}
\and M. Uemura\inst{11}
\and M. Villata\inst{14}
}

\offprints{G. Tagliaferri (gtagliaf@merate.mi.astro.it)}

\institute{Osservatorio Astronomico di Brera, Via Bianchi 46, I--23807 Merate, Italy
\and {\it Beppo}SAX Science Data Center, ASI, Via Corcolle, 19,
          I--00131 Roma, Italy
\and Dipartimento di Fisica, Universit\'a La Sapienza, P.le Aldo Moro 2, 00185  Roma, Italy
\and Istituto di Fisica Cosmica G.Occhialini, CNR, Via Bassini 15, I-20133 Milano, Italy
\and SISSA/ISAS, via Beirut 2-4, 34014 Trieste, Italy
\and Osservatorio Astronomico di Brera, Via Brera, 28, I-20121 Milano, Italy
\and Osservatorio Astronomico, Universit\`a di Perugia, Via A. Pascoli,
          I-06100 Perugia, Italy
\and Dipartimento di Scienze, Universit\`a dell'Insubria, Via Valleggio 11, I-22100
     Como, Italy
\and Foggy Bottom Observatory, Colgate University, 13 Oak Drive 13346, Hamilton NY,  USA
\and Dept. of Physics and Astronomy, Western Kentucky University, 1 Big Red Way, Bowling
     Green, KY 42102-3576,  USA
\and Dept. of Astronomy, Kyoto University, Kitashirakawa-Oiwake-cho, Sakio-ku, Kyoto 606-8502,
     Japan
\and Abastumani Astrophysical Observatory, 383762, Abastumani, Republic of Georgia
\and Dept. of Astronomy, Boston University, 725 Commonwealth Avenue, Boston MA 02215, USA
\and Osservatorio Astronomico di Torino, Strada Osservatorio 20, I-10025
     Pino Torinese, Italy
}

\date{Received 8 October 2001 / Accepted 7 December 2001}

\titlerunning{The X-ray spectrum of BL Lacertae}
\authorrunning{Ravasio et al.\ }

\abstract{
We report on two {\it Beppo}SAX observations of BL \,Lac (2200+420)
performed respectively in June and December 1999, as part of
a ToO program to monitor blazars in high states of activity.
During both runs the source has been detected up to 100 keV, but
it showed quite different spectra: 
in June it was concave with a very hard component above 5-6 keV
($\alpha_1 \sim 1.6$; $\alpha_2 \sim 0.15$); in December
it was well fitted by a single power law ($\alpha \sim 0.6$). 
During the first {\it Beppo}SAX observation BL \,Lac showed an 
astonishing variability episode: the $0.3 - 2$ keV flux doubled
in $\sim 20$ minutes, while the flux above 4 keV was almost contant.
This frequency--dependent event is one of the 
shortest ever recordered for BL\,Lac objects and places lower 
limits on the dimension and magnetic field of the emitting region 
and on the energy of the synchrotron radiating electrons.
A similar but less extreme behaviour is detected also in optical 
light curves, that display non-simultaneous, smaller fluctuations 
of $\sim 20 \%$ in 20 min. 
We fit the spectral energy distributions with a homogeneous, one-zone 
model to constrain the emission region in a very simple but effective
SSC + external Compton scenario, highlighting the importance of the 
location of the emitting region with respect to the Broad Line Region
and the relative spectral shape dependence. We compare our data with 
historical radio to $\gamma$-ray Spectral Energy Distributions.
}

\maketitle

\section{Introduction}

Blazar objects are highly variable sources characterised by 
non--thermal emission that dominates from the radio to the $\gamma$-rays.
This emission is supposedly due to a relativistic jet seen at a small 
angle to the line of sight (Blandford \& Rees 1978).
Although blazars emit over the entire electromagnetic spectrum,
their variability seems to be more pronounced in the optical-X-ray
band than the radio-infrared one, both in term of flux and time scale
(Ulrich et al. 1997).
It is also well known that the overall spectral energy distribution
(SED) of blazars shows (in a $\nu$ vs $\nu F_{\nu}$ representation)
two broad emission peaks; a lower frequency peak believed to be
produced by synchrotron emission and a higher frequency
peak probably due to the inverse Compton process. However,
during strong variability events, the overall SED can change
significantly. For instance, changes up to few orders of magnitude
in the position of the synchrotron peak have been detected
during flares in \object{Mkn\,501} (Pian et al. 1998) and
\object{1ES\,2344+514} (Giommi et al. 2000). Incidentally, 
both sources belong to the HBL class, High Frequency Peaked 
blazars, with the synchrotron peak in the UV -- X-ray band. 
A key to
understand blazar variability is the acquisition of wide
band spectra during major flaring episodes. Spectral and temporal
information greatly constrain the jet physics, since different
models predict different variability as a function of wavelength.

To this end, we started a project to observe
with the {\it Beppo}SAX satellite (Boella et al. 1997a)
blazars while they were in an
active state as detected in the optical, X-ray or TeV bands.
As part of this program, we already observed the two sources
\object{ON\,231} and \object{PKS\,2005-489} (Tagliaferri et al. 2000, 2001).
Here we present two {\it Beppo}SAX observations of the third 
observed source, \object{BL\,Lac} itself,
 carried out in June and December 1999, together with simultaneous optical 
and radio data. 
Other two observations have been carried out on the objects 
\object{OQ\,530} and \object{S5\,0716+714} (paper in preparation).

BL Lacertae, being the prototype of the BL \,Lac class, is one of 
the best--studied objects. It is a featureless LBL (low frequency 
peaked blazars) object, but sporadically it shows optical emission
lines (EW $\sim 6$ \AA; Vermeulen et al. 1995; Corbett et al. 2000). 
In the X-ray
it has been observed by many satellites; {\it Einstein}
detected a photon index of $1.68 \pm 0.18$ (Bregman et al. 1990),
Ginga a photon index in the range 1.7-2.2 (Kawai et al. 1991);
ROSAT a photon index of $1.95 \pm 0.45$ (Urry et al. 1996).
ASCA observed BL\,Lac in 1995 detecting a photon index of
$1.94 \pm 0.04$ (Madejski et al. 1999; Sambruna et al. 1999).
In July 1997 following an optical outburst, it was observed
with EGRET, {\it Rossi}XTE and ASCA. EGRET found that the flux level
above 100 MeV was 3.5 times higher than that observed in 1995
(Bloom et al. 1997). {\it Rossi}XTE found
a harder spectrum with a photon index in the range 1.4-1.6 over
a time span of 7 days (Madejski et al. 1999).
A fit to simultaneous ASCA and {\it Rossi}XTE data shows the existence 
of a very steep and varying soft component below 1 keV, photon index 
in the range 3-5, in addition to the hard power law component with
a  photon index of 1.2-1.4. Two rapid flares with time scales of
2-3 hours were detected by ASCA but only in the soft part of the 
spectrum (Tanihata et al. 2000). Finally, in November 1997, BL\,Lac
was observed with {\it Beppo}SAX that detected a photon index of
$1.89 \pm 0.12$ (Padovani et al. 2001).

The {\it Beppo}SAX observations presented here were triggered
when the source
was in very high optical states (R $\sim 12.5$), but when the
source was actually observed by {\it Beppo}SAX the optical flux
was lower (R = 13.4 - 13.6). In both observations the source was
clearly detected up to 100 keV giving us the possibility to study
BL\,Lac over an unprecedentedly large spectral range (0.3-100 keV)
with simultaneous data.

\section{X-ray observations}

\subsection{Observations and Data Reduction}

The {\it Beppo}SAX satellite carries on board four Narrow
Field Instruments (NFI) pointing in the same direction and covering a 
very large energy range from 0.1 to 300 keV (Boella et al. 1997a).
Two of the four instruments have imaging capability, the Low Energy
Concentrator Spectrometer (LECS), sensitive in the range 0.1--10 keV
(Parmar et al. 1997), and the three Medium Energy Concentrator
Spectrometers  (MECS) sensitive in the range 1.3--10 keV (Boella
et al. 1997b). Only two MECS detectors were functioning
 at the time of our observation. The LECS and  MECS detectors are all Gas
Scintillation Proportional Counters and are at the focus of identical
grazing incidence X--ray telescopes. The other two are passively collimated
instruments: the High Pressure Proportional Counter (HPGSPC), sensitive
in the range 4--120 keV (Manzo et al. 1997) and the Phoswich Detector
Systems (PDS), sensitive in the range 13--300 keV (Frontera et al. 1997). 
For a full description of the {\it Beppo}SAX mission see 
Boella et al. (1997a).

\begin{table*}
\begin{center}
\begin{tabular}{|l|cccccc|}
\hline
date & LECS  & count rate$^{\rm a}$ & MECS & count rate$^{\rm b}$ & 
PDS & count rate$^{\rm c}$ \\
& exposure (s) & $\times 10^{-2}$ cts \ s$^{-1}$ & exposure (s) & 
$\times 10^{-2}$ cts \ s$^{-1}$ & 
exposure (s) & cts \ s$^{-1}$ \\
\hline
& & & & & &\\
5-7 June 1999     &  45253 & $7.14 \pm 0.14$  & 54404 & $11.1 \pm 0.2$  & $3.9\times10^4$ & 
    $0.16 \pm 0.02$ \\ 
5-6 December 1999 &  17543 & $8.81 \pm 0.25$  & 54677 & $17.8 \pm 0.19$ & $2.4\times10^4$ &
    $0.18 \pm 0.03$ \\
\hline \hline
 & B & V & R$_{\rm c}$ & I$_{\rm c}$ & $\alpha_{opt}$ $^{\rm d}$ & \\
\hline
& & & \hspace{-1.2cm} magnitudes & & & \\
\hline
06 June 1999     & $15.01 \pm 0.03$ & $14.13 \pm 0.03$ & 
      $13.52 \pm 0.03$ & $12.79 \pm 0.03$ & $1.35 \pm 0.06$ & \\
06 December 1999 & $15.33 \pm 0.03$ & $14.26 \pm 0.03$ & 
      $13.49 \pm 0.03$ & $12.69 \pm 0.03$ & $2.02 \pm 0.06$ & \\
\hline
\end{tabular}
\label{table1}
\end{center}
\caption{Journal of observations.
$^{\rm a}$ In the band 0.1--10 keV;
$^{\rm b}$ for two MECS units in the band 1.5--10 keV; 
$^{\rm c}$ in the band 12--100 keV;     
$^{\rm d}$ calculated from mJy fluxes. Optical data were dereddened using
A$_V = 1.09$.}
\end{table*}

The log of the BL\,Lac observations is given in Table 1, together with
the exposure times and the mean count rates in the various instruments.
 The data
analysis for the LECS and MECS instruments was based on the linearized,
cleaned event files obtained from the online archive (Giommi \& Fiore 1998). 
Light curves and spectra were accumulated with the FTOOLS package (v. 4.0),
using an extraction 
region of 8 and 4\,arcmin radius for the LECS and MECS, respectively.
The LECS and MECS background is low and rather stable, but not uniformly 
distributed across the detectors. 
For this reason, it is better to evaluate the
background from blank fields, rather than in concentric rings around the 
source region.
Thus, after having checked that the background was not varying during the 
whole observation by analyzing a light curve extracted from a source--free 
region, we used for the spectral analysis the background files accumulated
from long blank field exposures and available from the SDC public ftp site
(see Fiore et al. 1999, Parmar et al. 1999).

The PDS was operated in the customary collimator rocking mode, where half
the collimator points at the source and half at the background and they are
switched every 96\,s. The data were analysed using the XAS software
(Chiappetti \& Dal Fiume 1997) and the data reduction was performed according
to the procedure described in Chiappetti et al. (1999), inclusive of spike
filtering.

The source was not detected by the HPGSPC detector.

\subsection{Spectral analysis}

\begin{figure}
\centerline{
\vbox{
\psfig{figure=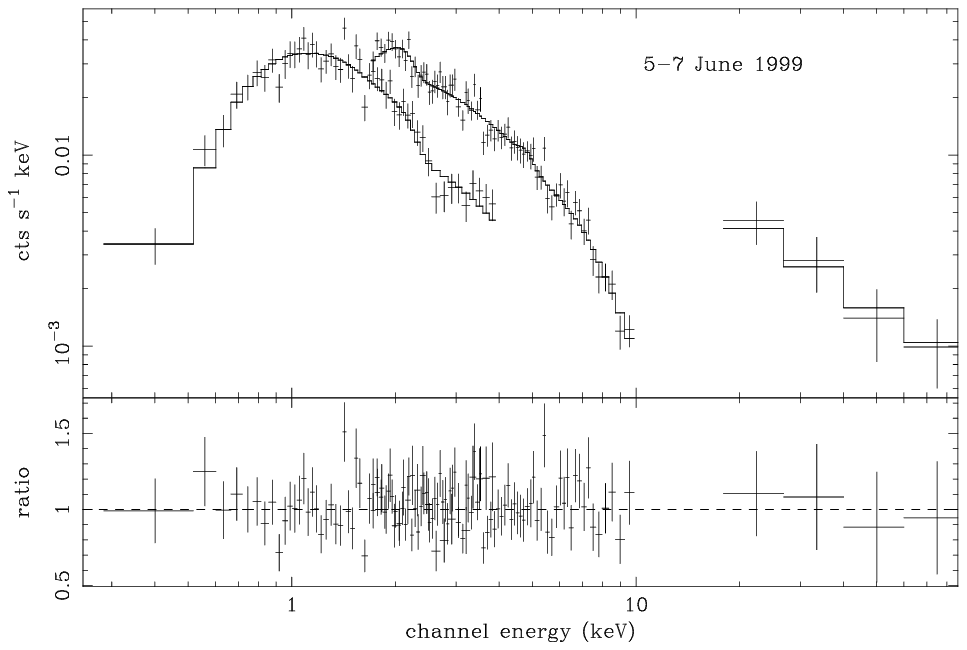,width=9.5cm}
\vskip0.5cm
\psfig{figure=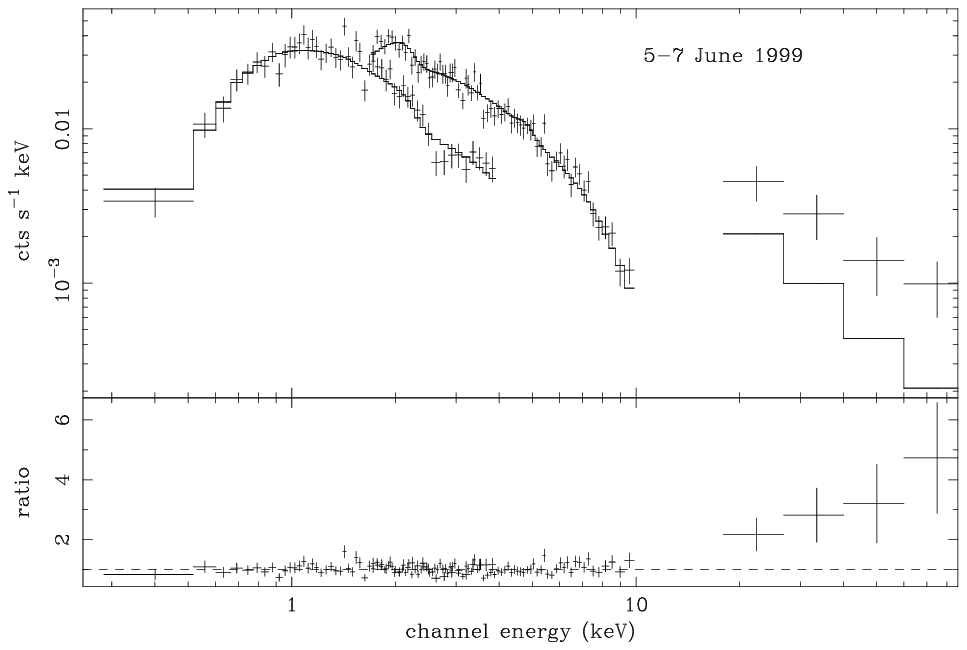,width=9.5cm}
}}
\caption{Top panel: LECS+MECS+PDS \object{BL\,Lac} spectrum during
the June
1999 observation, the best fit model is a sum of two power laws.
Bottom panel: the same spectrum fitted with a power law model
(over the entire energy range), clearly showing that a simple
power law model can not fit the data.}
\label{june-spec}
\end{figure}

The spectral analysis was performed with the XSPEC 10.0 package,
fitting together the LECS, MECS and PDS data. To fit the data of 
these detectors simultaneously, one has to allow for a constant
rescaling factor to account for intercalibration systematics
of the instruments. The acceptable values for these constants are in
the range 0.67-1 for the LECS and in the range 0.77-0.93 for the
PDS, with respect to the MECS (Fiore et al. 1999). In order to
reduce the number of fitting parameters, during the parameter error
search we fixed the two constants at the best-fit values of 0.72 and
0.9 in all our fits, respectively. The LECS data have been considered
only in the 0.1--4 keV interval, as suggested by SDC team 
(Fiore et al., 1999).
We first considered the June observation.
We fitted the data
with a single power law plus free interstellar absorption. The 0.3-100 
keV X-ray spectrum of June is not well fitted by this model,
showing large residuals above 10 keV (see fig. \ref{june-spec} 
bottom panel and Table\,2) and best fit estimate of 
N$_{\rm H}$ significantly lower than the Galactic interstellar
value of N$_{\rm H} = 2.02 \pm 0.01 \times 10^{21}$ cm$^{-2}$
(Elvis et al. 1989). A good fit to the data is obtained using a sum
of two power laws (see fig. \ref{june-spec} top panel and Table\,2).
The energy intersection falls between 5 and 6 keV. Below we 
detect a softer component ($\alpha = 1.6 \pm 0.25$),
 probably due to synchrotron emission, 
while above we observe a harder component ($\alpha = 0.15 \pm0.22$),
 supposedly originating from
inverse Compton emission. The interstellar absorption is higher than
the Galactic one and is consistent with
previous results (Sambruna et al., 1999; Madejski et al., 1999). This
X-ray spectrum is very similar to the one observed for ON\,231, while
it was in an active state (Tagliaferri et al. 2000). A break in the
combined ASCA and {\it Rossi}XTE 0.7-20 keV spectrum was also detected
during a short flaring state of BL\,Lac in July 1997
(the {\it Rossi}XTE exposure time was of about 30 minutes).
 However, besides the source being in a higher state, the 
break was at about 1 keV and the synchrotron spectrum was much steeper,
with an energy index $\alpha = 2.7$ (Tanihata et al. 2000). Thus,
although in both cases the synchrotron and the Compton components
were simultaneously detected, the X-ray spectrum of BL\,Lac was quite 
different. During the July 1997 outburst the Compton component was brighter,
 while during our June 1999 observation the synchrotron component 
 increased its contribution to the overall X-ray spectrum,
dominating up to $\sim 5$ keV.

After the June 1999 {\it Beppo}SAX observation, the source remained active
 in the optical,
with the R magnitude ranging between $\sim 12.6 - 13.8$ mag. At the end
of November it brightened again to R$\sim 12.4$, thus we triggered a second
{\it Beppo}SAX ToO observation. However, when the source was observed
in the X-rays the R magnitude was down to R$\sim 13.5$.
During this second observation, 
the X-ray spectrum was in a quite different state. In this
case the 0.3-100 keV spectrum is well fitted by a single
power law model (see Table\,2). The total flux in the 2-10 keV energy band 
is about a factor of two higher than in June.
However, the two spectra cross at about 1.5 keV, thus the flux
in the softer X-ray energy band is actually lower.
Clearly, the overall shape of the SED changed between the two observations
and during the second one we detected mostly the Compton component.
The 2-10 keV flux is very similar to the value detected during the June 1997
{\it Beppo}SAX observation, which was also well fitted by a single power
law that is a little bit steeper, $\alpha = 0.89 \pm 0.12$,
 (Padovani et al. 1998, 2001).

\begin{table*}
\begin{tabular}{ccccccccc}
\hline
Date & N$_{\rm H}$ & $\alpha_1$ & $\alpha_2$ & $F_{1keV}$ & $F_{0.5-2 keV}$ & $ F_{2-10 keV}$ & $F_{10-100 keV}$ & $\chi_r^2/d.o.f.$\\
 & $\times 10^{21} \ {\rm cm^{-2}}$ & & & $\mu$Jy & ergs cm$^{-2}$ s$^{-1}$ & ergs cm$^{-2}$ s$^{-1}$ & erg cm$^{-2}$ s$^{-1}$ & \\
\hline
5-7 June & $1.41 \pm 0.02$ & $1.04 \pm 0.03$ & & 1.77  & $0.4 \times10^{-11}$ & $0.6\times10^{-11}$ & $0.86\times10^{-11}$ & $1.2/124$\\
\hline
5-7 June$^1$ & $2.42 \pm 0.04$ & $1.57 \pm 0.25$ & $0.15 \pm 0.22$ & $0.77$
& $0.4\times10^{-11}$ & $0.6 \times 10^{-11}$ & $2.48\times10^{-11}$ & $1.03/122$\\
\hline
5-6 Dec. & $2.49 \pm 0.05$ & $0.63 \pm 0.06$ & & $0.96$ & $0.35\times10^{-11}$& $1.2 \times 10^{-11}$  & $3.67\times10^{-11}$ & $ 1.02/130$\\ 
\hline
\end{tabular}
\caption{
$^1$The two power laws cross at $\sim 5.5$ keV. }
\label{table2}
\end{table*}

The interstellar absorption, $N_{\rm H}$, plays an important role
in fitting the X-ray spectrum of BL\,Lac, due to its high value and
to the presence of a molecular cloud in the line of sight 
(Lucas \& Liszt, 1993).
 To determine the best $N_{\rm H}$ value to use to fit the 
X-ray spectrum of BL\,Lac, we repeated the fitting process fixing
the $N_{\rm H}$ both to the Galactic column density of neutral hydrogen
from 21 cm measurements
$N_{\rm H} = 2 \times 10^{21}$ cm$^{-2}$ (Elvis et al., 1989) 
and to the higher value $N_H = 3.6 \times 10^{21}$ cm$^{-2}$
obtained taking also to consideration the absorption due to the
presence of the molecular cloud.
In the first case, we cannot find a good fit to the data for
both observations. The second  value gave an acceptable fit for 
the first observation ($\chi_r^2/d.o.f.= 1.06/123$): an F-test 
suggested that letting $N_H$ remain free does not produce a significant 
improvement in the quality of the fit.
This is not true for the December spectrum: for this observation,
letting $N_H$ remain free gives a probability $\geq 95 \%$ of improving the 
quality of the fit. Thus, we kept the $N_H$ free to vary in both
observations, finding values that are perfectly consistent.
A similar value was also found by Padovani et al. (2001) for the first
{\it Beppo}SAX observation. Given that the LECS is well suited for
the determination of the interstellar absorption in the X-ray band,
we think that the value found in three different {\it Beppo}SAX 
observations (and with different best fit models) should be the one
used for the analysis of the BL\,Lac X-ray data. Thus, the value 
used for the analysis of the ASCA data is probably too high (Sambruna
et al. 1999, Tanihata et al. 2000).

\begin{figure}
\centerline{
\vbox{
\psfig{figure=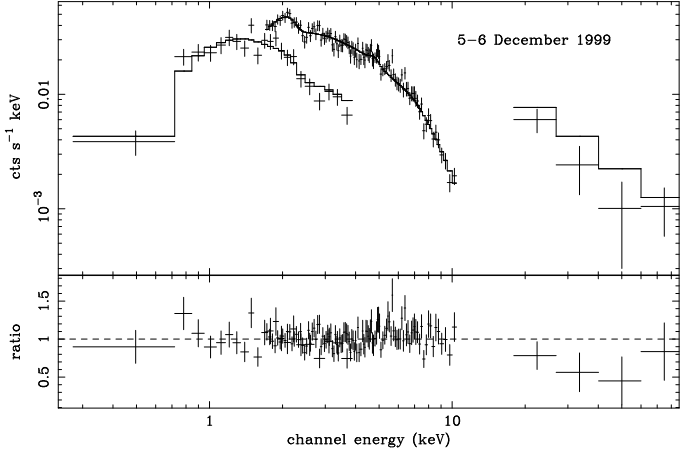,width=9.5cm}
}}
\caption{In the December 1999 observation the data are well fitted by
a single power law model plus an interstellar absorption value perfectly 
consistent with the value obtained for June's spectrum.}
\label{pow-dec}
\end{figure}

\subsection{Temporal analysis}

\begin{figure*}
\centerline{
\vbox{
\psfig{figure=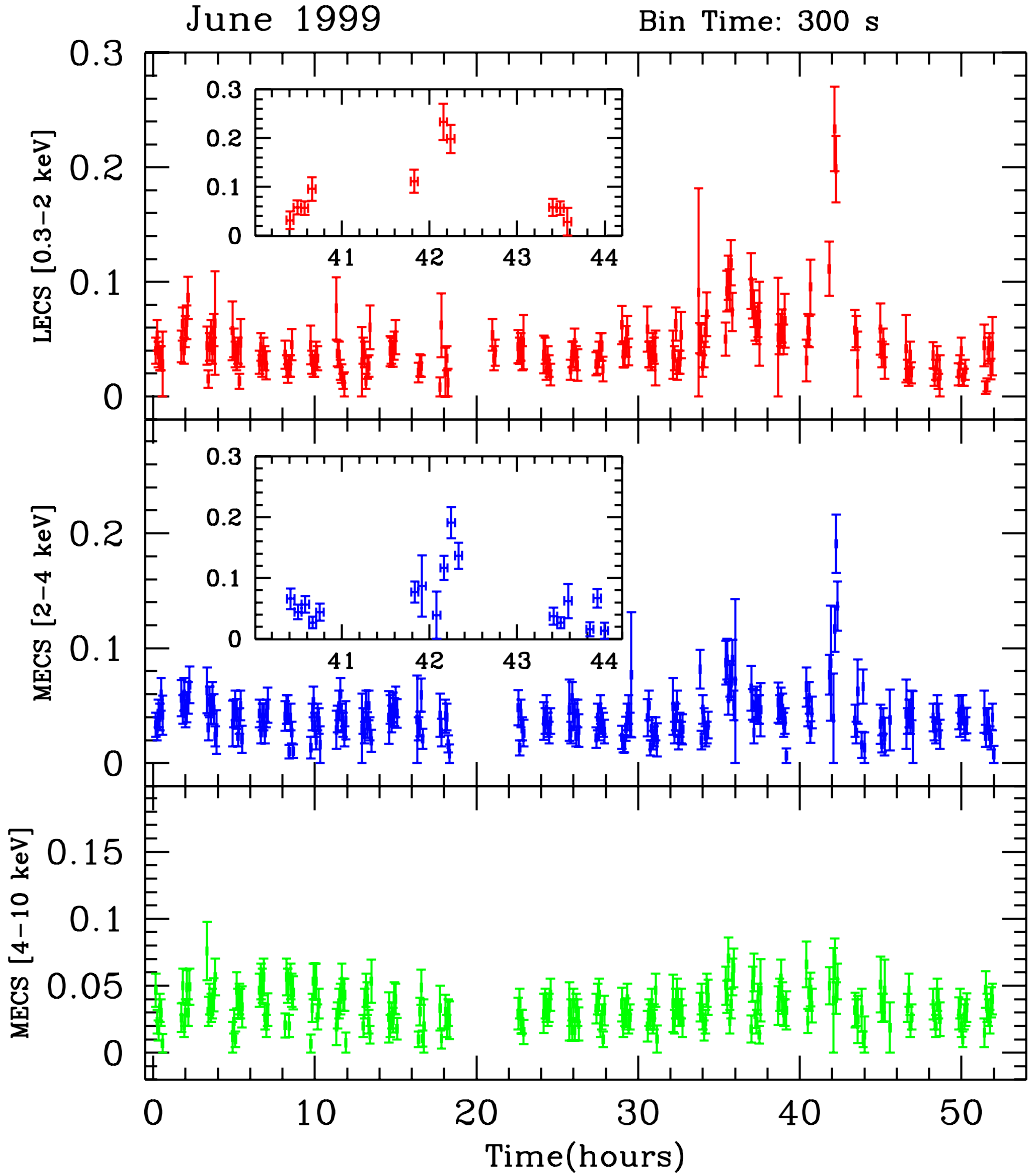,width=16cm}
}}
\caption{LECS 0.3-2 keV (top panel), MECS 2-4 keV (mid panel) and
MECS 4-10 keV (bottom panel) light curves of BL\,Lac during the June 
1999 {\it Beppo}SAX observation. Note the clear flare detected at the 
end of the observations by different detectors (top and mid panels)
and better shown in the insert. Note also that above the spectral break,
i.e. in the Compton component, the flare is clearly not present
(bottom panel).}
\label{curva-june}
\end{figure*}

The X-ray light curve of the June observation looks in general
quite constant. However, it reveals an amazing feature 
about 42 hours after the beginning of the measurements. 
In a very short timescale, about 20 minutes, the [0.3-2 keV] flux doubled:
this is the fastest variability ever measured for BL\,Lac.
In 2 hours the soft X flux increased 4 times and then decreased 
to former values. 
This is not frequency--independent: as can be seen in Fig. \ref{curva-june},
variations are detected only in the softer spectral component.
While the synchrotron emission is extremely variable, the inverse Compton 
component remains almost constant during the observation. This behaviour
is very similar to that observed for S5 0716+714
(Giommi et al. 1999) and ON\,231 during a high state (Tagliaferri
et al. 2000). Also in these cases fast variability
was detected only for the synchrotron part of the spectrum .
A similar behaviour for BL\,Lac has been observed with ASCA
during the July 1997 campaign, when the source was in a high state. Again,
rapid variability was detected only in the soft part of the X-ray 
spectrum, with the 0.7-1.5 keV flux flaring a couple of times
by a factor of two on time scales of 3-4 hours, while
above 3 keV no rapid variability was found (Tanihata et al. 2000).
In analogy with the {\it Beppo}SAX observations of ON\,231 and BL\,Lac
one would expect that also in the ASCA data there is a break between 1-3
keV, reflecting the presence of the synchrotron tail responsible for
the fast variability below 1.5 keV and of the less variable Compton
component above 3 keV. However, the presence of this break is not so clear
in the ASCA data and only in a simultaneous {\it Rossi}XTE and ASCA
spectrum  can one see that the spectrum is much steeper below 1 keV
(Tanihata et al. 2000).
Once again, the {\it Beppo}SAX large spectral energy range combined
with its good sensitivity and energy resolution is unique in
determining the X-ray spectral shape of bright blazars in the range
0.1-100 keV.

The light curves of the December observation do not reveal variability,
but this is not surprising as {\it Beppo}SAX detected only the inverse
Compton component.

\section{Optical observations}

During the {\it Beppo}SAX observations BL\,Lac was intensively monitored in
the optical by several observers (see Table 3) located at different longitudes
around the world, belonging to the Whole Earth Blazar Telescope (WEBT) 
collaboration (Villata et al. 2000). This collaboration, begun in 1997, 
aims to provide 24 hour coverage of Blazars.

\begin{table}[b]
\begin {tabular}{l l l l}
\hline
 Group               & Bands  &Tel. & detector \\
\hline
 Abastumani        & B,V,R,I &    70 cm  & Texas TC241   \\
 Lowell Obs.       & V, I    &   180 cm  & SITe 501A     \\
 Foggy Bottom      & R,I,V   &    40 cm  & CCD           \\
 Kyoto             & R       &    25 cm  & Kodak KAF 400 \\
 Perugia           & R, I    &    40 cm  & Texas TC211   \\
 Roma I            & V       &    50 cm  & Texas TC241   \\
 Roma II           & I       &    35 cm  & SITe 501A     \\
 Torino            & B,V,R,I &   105 cm  & Astromed EEV CCD \\
\hline
\end {tabular}
\caption{WEBT optical telescopes participating in the BL Lac
campaign. Roma I is located in Vallinfreda, Roma II in Greve}
\end{table}

Each group performed differential photometry of BL\,Lac with respect
to 4 nearby reference stars (Bertaud et al. 1969; Fiorucci \& Tosti 1996).
The data were then checked for systematic effects and finally a
self-consistent light curve of BL\,Lac was obtained.

Not all the involved observers used the same filters (see
Table 3), nor made observations with all the filters.
To build an optical light curve as complete as possible
in time coverage we transformed observations in different bands into the
I band, using 
appropriate colour indices for each Observatory.
A small change of the colour index with the source luminosity (0.06 in V-I
for a change of 0.2 R mag) was clearly detected in the Abastumani data set,
but it does not  seriously  affect
the transformation of the observed magnitudes from 
other bands into the I band, so that the long-term behaviour of the source 
during the run can be safely traced. 
The resulting light curve for the June pointing
is plotted in Fig. 4 (lower panel) together
with the X-ray one (upper panel) for comparison.

It is apparent from Fig. 4 that BL\,Lac varied continuously during the
{\it Beppo}SAX pointing with an overall amplitude of 0.55 mag.
 After the maximum
at I=12.45 just before the beginning of the {\it Beppo}SAX observation, 
followed by a decay of about 0.4 mag in less than 10 hours,
the source behaviour was characterised by an increasing trend for about
one day, with small oscillations; the brightness then
decreased, with a short plateau, down to I = 13.

The comparison with the X-ray light curve is not easy, due to the different
time sampling and accuracy of the two curves. In any case the amplitude of
the X-ray variability is much wider than the optical one. 
The most striking feature of
the X-ray light curve, i.e. the strong and fast flare discussed above, has
no optical counterpart, while a temporal coincidence may be found between
the smaller X-ray flare at JD 1336.3 and the fast decrease in the optical.
However this cannot be stated for sure, due to the X-ray error bars and the
lack of continuity in the optical light curve. It is interesting to note
that this optical event
occurred about 6 hours before the stronger soft X-ray flare, that does not
seem to have an optical counterpart. Thus, it seems that there is not 
a one--to--one connection between fast X-ray and optical variability.

During the second run (December) the weather was not so good and
the coverage from the ground was less complete. The average flux
level in the I band was 12.6, about the same as in June,
while the (V-I) colour index was substantially steeper ($\sim 1.6$).
The source showed a monotonic decrease of 0.2 mag at the beginning of the
observing run and then remained nearly constant. Linear polarization
was measured on December 14, 1999, at a level of $4\%$ (Tommasi et al., 2001).

Simultaneous values of magnitudes for the June and December pointing are
given in Table 1, together with the derived spectral slopes.
The steep spectral slope, when extrapolated to the soft X-rays, predicts a
flux below the observed one, confirming the interpretation of the observed
X-ray spectrum as due to the Inverse Compton component.
This behaviour is in agreement with the X-ray light curve, which remained
constant during the second pointing.

\begin{figure*}
\centerline{
\vbox{
\psfig{figure=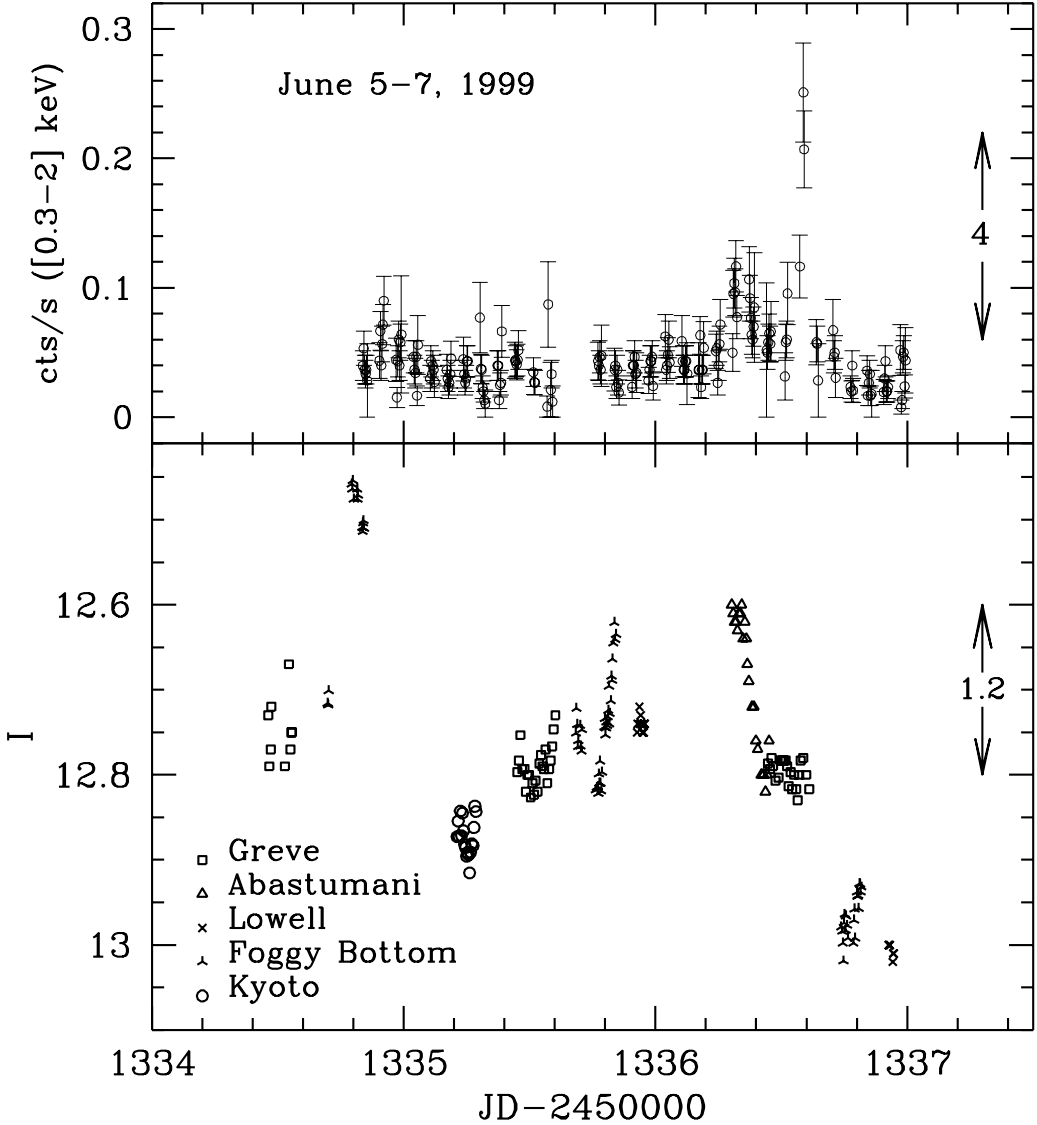,width=16cm}
}}
\caption{Soft X-ray and optical (I band) light curves of BL\,Lac
during 1999 June, 5-7. The scale of variability is also reported
in the two panels.}
\label{ottix-j}
\end{figure*}

\section{Discussion}

During our two {\it Beppo}SAX observations, BL\,Lac showed two different
 spectral shapes not only in the X ray band, but also in the optical one.
Even if optical fluxes are almost similar, the spectrum in June is
harder than in December. Optical magnitudes, fluxes and spectral
indices are listed in Table\,1. 
The synchrotron spectrum of December seems to be shifted 
toward lower energies than that of June: its optical spectrum is softer
and {\it Beppo}SAX detected just the inverse Compton component. 
The observed Compton spectra are quite different, as we already mentioned, 
being much harder in June ($\alpha_C = 0.15$) than in December 
($\alpha_C = 0.63$), and even more so with respect
to the previous {\it Beppo}SAX observation 
($\alpha_C= 0.9$, Padovani et al. 2001).  In a simple picture one would have
expected that the presence of a break in the X-ray spectrum, and hence 
of a synchrotron component, would have implied a higher source flux. 
But this is not the case.
Although the flux below 1.5 keV was higher during the observation 
with the spectral break, the 2-10 keV flux was higher 
in the other two {\it Beppo}SAX observations.
 All this can be explained by the fact that both the synchrotron, 
as shown by the high optical state, and the Compton 
components are varying.  In June, 
the synchrotron flux prevails in the soft X-ray band and two 
components are detected in the X-ray spectrum. However, the 2-10 keV band is 
still dominated by the Compton emission, which is lower but 
unusually hard.

\begin{figure*}
\centerline{
\vbox{
\psfig{figure=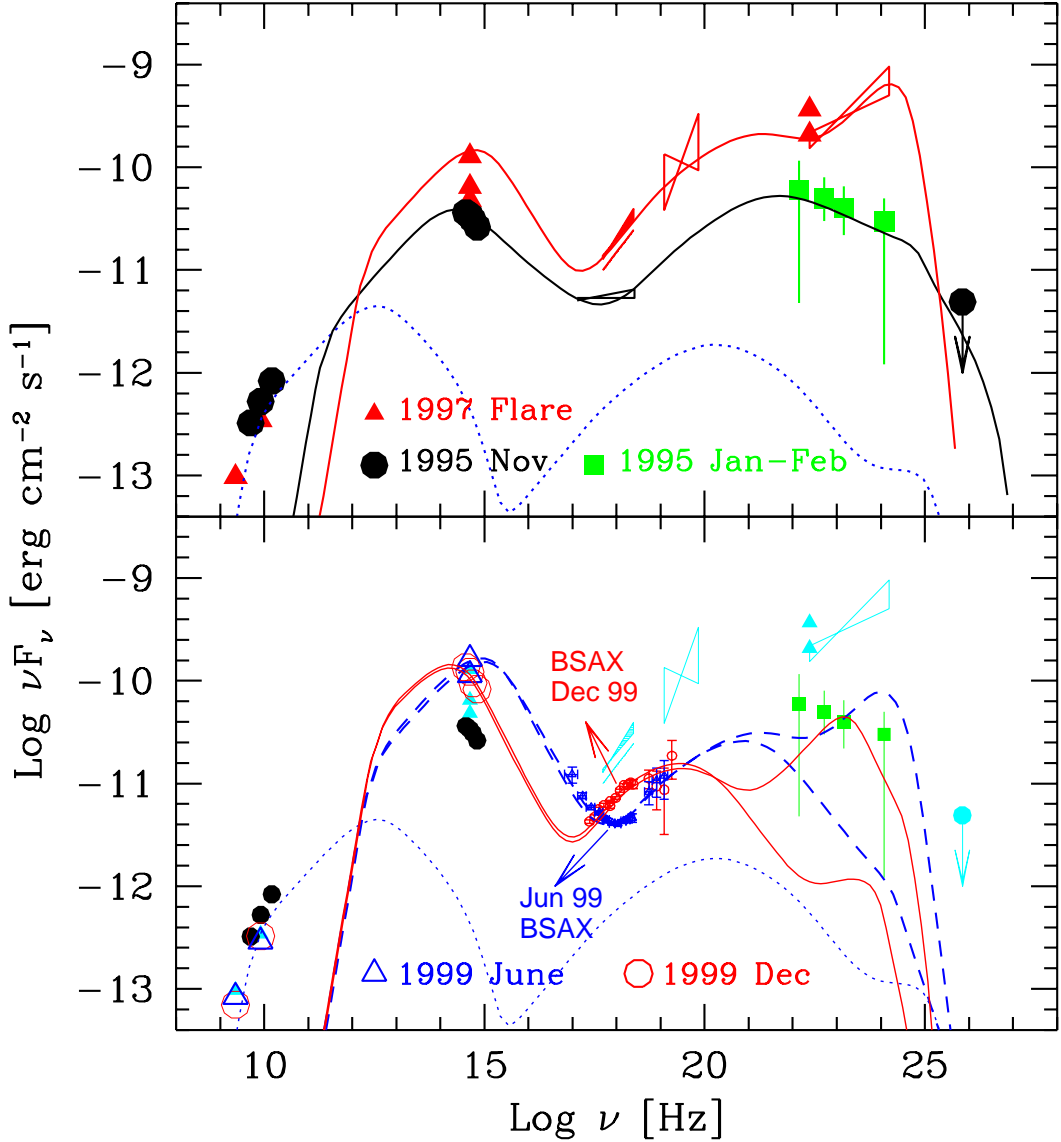,width=16cm}
}}
\caption{Four simultaneous SEDs of BL\,Lac together with our
best fit models. For the 1995 SED see Sambruna et al. (1999), for the
1997 SED see  Madejski et al. (1999). For the two {\it Beppo}SAX
observations the simultaneous radio data are from NRAO, while the
optical data are from our observations. Note that the EGRET data 
of 1995 are not simultaneous with the other data of the same year.
Top panel: the 1997 data modeled with a synchrotron inverse 
Compton model where the emission line photons are important for the 
formation of the very high energy spectrum. 
Note however that the emission in the entire X--ray band is produced
by the synchrotron self--Compton process.
For the 1995 SED, the fainter and steeper EGRET spectrum suggests that the 
emission line photons are not important. Consequently we have assumed that 
the emitting region is located beyond the BLR.
Bottom panel: triangles and circles correspond to the SEDs of June and December 1999,
as labeled. For comparison we also report the 1995 and 1997 SEDs (crosses).
For each SEDs, we show the models corresponding to important or negligible
emission line radiation for the formation of the $\gamma$--ray spectrum.
For comparison, we also plot the 1995 and 1997 data in light grey.
In both panels we also show the calculated emission produced in a much larger
region of the jet, to account for the radio flux (dotted line).
}
\label{seds}
\end{figure*}

To show the complex behaviour of the spectral emission of BL\,Lac
we plot in Fig. \ref{seds} four different SEDs corresponding to 
the multi-wavelength campaigns carried out during November 1995,
during July 1997, when the source was in a very high state and 
during our two {\it Beppo}SAX observations. This figure clearly shows 
the high degree of variability and complexity of BL\,Lac's SED.

\subsection{Limits on magnetic field and particle energies}

During the {\it Beppo}SAX observation of June 1999, the X--ray 
flux below 4 keV varied by a factor 2 in 20 minutes (see Fig. 3).
This is one of the fastest variability episodes ever observed at any frequency 
in a BL \,Lac object, to be compared with a very similar timescale 
observed in the TeV band in Mkn 421 (Gaidos et al. 1996;
see also Maraschi et al. 1999 and Fossati et al. 2000 for the 
correlation between the X--ray and TeV emission).
Interestingly, the higher energy X--rays, belonging to the flat 
part of the observed spectrum, did not vary.
This suggests that, in the two X--ray bands, we see the emission by a different 
part of the electron population: the highest energy electrons, that
form the steep tail of the synchrotron spectrum, radiatively cool more rapidly
than electrons producing the inverse Compton part of the spectrum
visible above $\sim 5$ keV.
This implies that a similarly fast variability should be observed in the tail
of the Compton spectrum (as indeed was the case for Mkn 421).

The size of the region emitting the variable X--rays 
must be $R<ct_{var}\delta$, where 
$\delta\equiv [\Gamma(1-\beta\cos\theta)]^{-1}$ is the beaming factor.
$\Gamma$ and $\beta$c  are the bulk Lorentz factor and the bulk velocity,
respectively, and $\theta$ is the viewing angle.

We then require that the cooling time (in the comoving frame)
is shorter than $R/c$ for those electron emitting at $\nu =10^{17}\nu_{17}$ Hz.
This in turn implies a lower limit on the magnetic field and an upper limit
on the energy of the electrons emitting at $\nu$:
\begin{equation}
B\, >\, 1.2 \, t_h^{-2/3} (1+U_B/U_r)^{-2/3} \nu_{17}^{-1/3} \delta^{-1/3} (1+z)^{1/3}
\,\, {\rm G}
\end{equation}
\begin{equation}
\gamma\, <\, 1.5\times 10^5 \, t_h^{1/3} (1+U_B/U_r)^{1/3} \nu_{17}^{2/3} \delta^{-1/3}
(1+z)^{1/3}
\end{equation}
where $t_h$ is the variability timescale in hours and
$U_B/U_r$ is the magnetic to radiation energy density ratio
(we implicitly assume that Klein Nishina effects are negligible).
Using $t_h=1/3$, $U_B/U_r=1$ and $\nu_{17}=1$ we derive 
$B>1.6\delta^{-1/3}$ G and
$\gamma< 1.3\times 10^5 \delta^{-1/3}$.

More severe limits can be set by the optical variations,
in the $I$ band ($\nu\sim 4\times 10^{14}$ Hz) of about 20 per cent
in 20 minutes. 
Assuming a doubling time of 1.5 hours, we obtain
$B>3.7\delta^{-1/3}$ G and 
$\gamma< 5.5\times 10^3 \delta^{-1/3}$.

Above 4 keV, the radiation is due to the inverse Compton process,
and are characterized by smaller energies than those
producing the synchrotron tail.
Their random Lorentz factor can be estimated assuming that the 
relevant seed photons have frequencies $\sim 10^{14}$ Hz
(i.e. the synchrotron peak), resulting in $\gamma\sim 100$.
Assuming $B>3.7 \delta^{-1/3}$ G, their cooling time 
(in our observing frame) is less than $\sim 78 \delta^{-1/3}$ hours,
which comfortably allows for the non--detection of variability 
in the hard X--ray band.
Longer timescale are derived in the case of scattering with 
photons of the BLR, since in this case the relevant electron energies
producing photon in the hard X--ray band are smaller.

\subsection{Homogeneous synchrotron inverse Compton models}

We have applied a homogeneous, one--zone synchrotron inverse Compton 
model to the four SEDs shown in Fig. 5 (Ghisellini et al. 199).
The model assumes that the source is spherical, of radius $R$,
magnetic field $B$
and it moves with a bulk Lorentz factor $\Gamma$.
Radiation is beamed with the Doppler factor  $\delta$.
Throughout the source we inject relativistic particles 
continuously, at the rate $Q(\gamma)\propto \gamma^{-s}$ cm$^{-3}$
between $\gamma_{min}$ and $\gamma_{max}$.
We derive the stationary particle distribution solving the
continuity equation, taking into account radiative cooling, 
electron-positron pair production, and approximating the
Klein--Nishina cross section, for the scattering process,
by a step function, equal to the Thomson cross section 
if the seed photon has energy $h\nu \le m_e c^2/\gamma$ and zero otherwise.

In the absence of pair production and if the Klein--Nishina effects
are negligible, the particle distribution is a broken power law:
$N(\gamma)\propto \gamma^{-2}$ below $\gamma_{min}$
and $N(\gamma)\propto \gamma^{-(s+1)}$ above.

The seed soft photon distribution is provided by the synchroton process
only if the source is located at a distance, along the jet,
greater than a critical distance $z_{BLR}$ where most of the line
emission is produced.
Otherwise, we include the radiation energy density of the emission
lines, as seen in the frame comoving with the source.

The applied model is aimed at reproducing the spectrum 
originating in a limited part of the jet, thought to
be responsible of most of the emission. 
This region is necessarily compact, since it must account
for the fast variability.
Therefore the radio emission from this compact regions
is strongly self--absorbed, and the model cannot account
for the observed radio flux.
We have then calculated the spectrum produced in a much larger region
of the jet (with dimension $R=7\times 10^{17}$ cm), to account for the radio 
flux. 
This spectrum is shown in Fig. 5 with the dotted line, and the input parameters
are listed in Table 4.

The observed variability timescale $t_{var}\sim 20$ minutes severely 
limits the dimension of the compact source to be 
$R<c t_{var} \delta\sim 3.6\times 10^{13}\delta$ cm.
We have assumed $\delta=20$ and $R=7\times 10^{14}$ cm
for the SEDs of the 1997 flare and for the two SEDs observed
by us in 1999.
For the 1995 data, instead, we have assumed a  dimension 10 times larger.

This choice is suggested by comparing the overall SEDs of 1995 and 1997:
as can be noted, there is a large difference in the emission at high
energies above 100 MeV, which has been interpreted (Sambruna et al. 1999;
B\"ottcher \& Bloom 2000, Madejski et al. 1999) as a different contribution of the
external seed photons for the formation of the Compton spectrum.
In 1995, the steep and relatively weak $\gamma$--ray emission
favours SSC as the main emission process, while the flat
and strong $\gamma$--ray emission during the 1997 flare suggests
that external seed photons are relevant.

BL\,Lac has indeed shown variations in the intensity of the emission lines,
which have sometimes exceeded the canonical threshold of 5 \AA \ 
of equivalent width to define an objects as belonging to the BL \,Lac class
(Vermeulen et al. 1995).
However, the observed variation are within a factor of 3 in absolute intensity,
and it is not clear if this is enough to influence  strongly the production
of the high energy emission.

\begin{table*}
\begin{center}
\begin{tabular}{llllllllll}
\hline
Date      &In/Out &$L^\prime$ &$R$  &$B$ &$U^\prime_{ext}$ &$\delta$ &$s$ &$\gamma_{min}$ 
      &$\gamma_{max}$\\
          &     &erg s$^{-1}$ & cm  &G   &erg cm$^{-3}$    &         &     &              &  \\
\hline
1995      &Out  &9.0e40       &7e15 &0.2 &...              &20        &2.8 &3.5e3        &5.0e5   \\
1997      &In   &5.2e41       &7e14 &4.0 &1                &20        &3.8 &1.5e3        &3.0e4   \\
1999 Jun  &Out  &1.3e41       &7e14 &1.2 &...              &20        &3.7 &1.1e3        &1.0e5   \\
1999 Jun  &In   &1.6e41       &7e14 &1.1 &1                &20        &3.7 &1.2e3        &5.0e4   \\
1999 Dec  &Out  &1.0e41       &7e14 &13  &...              &20        &3.7 &4.7e2        &1.0e4   \\
1999 Dec  &In   &1.1e41       &7e14 &13  &1                &20        &3.8 &4.7e2        &1.0e4   \\
Large     &Out  &5.2e39       &7e17 &1.2e-3  &...          &20        &2.8 &6.0e3        &1.0e5   \\
\hline 
\end{tabular}
\end{center}
\caption{Input parameters for the SSC models}
\end{table*}

We propose here an alternative scenario, based on the idea that the
dissipation region of the jet corresponds to the collision of two 
blobs or shells moving at slightly different velocities
(the ``internal shock" scenario for blazars, see Ghisellini 1999;
Spada et al. 2001, Madejski et al. 1999).
If the bulk Lorentz factor of the two shells differs by a factor 2,
a later but faster shell catches up with an earlier and slower one
at a distance $z\sim \Gamma^2 z_0$, where $z_0$ is the initial separation 
and $\Gamma$ is the Lorentz factor of the slower shell.
We therefore have that different SEDs correspond to different collisions,
that can occur at different location in the jet. 
Some of them may be located within the size of the BLR, while
others occur outside.
This has a quite dramatic effect on the spectrum, since within the BLR the
energy density of the emission line radiation (as measured in the comoving frame)
easily exceeds the magnetic and the synchrotron energy densities.
Therefore, collisions occurring at $z<z_{BLR}$ emit 
most of their power in the GeV band, as happened in the 1997 flare.
On the contrary, for collisions beyond $z_{BLR}$ the emission
line radiation is negligible, and the Compton to synchrotron luminosity
ratio is controlled by the ratio between the magnetic field and the 
synchrotron energy densities.

We can estimate the size of the BLR of BL\,Lac by adopting the correlation
presented by Kaspi et al. (2000) between the ionizing luminosity
and the size of the BLR.
We derive $z_{BLR} \sim$3--5$\times 10^{16}$ cm.

For the two SEDs observed by us in 1999, we unfortunately lack the high
energy EGRET data, which best discriminate between collisions
occurring inside/outside the BLR.
In fact, the medium--hard energy X--rays are always produced by the
SSC process, even in the flare of the summer 1997, and therefore
the X--ray flux level and spectrum do not reflect the importance of the
external seed photons.

For the 1999 June SED, the very short variability timescale
suggests a very compact region, and hence a location, along the jet,
close to the jet apex (a transverse dimension $R=7\times 10^{14}$ cm
translates into a radial distance $z\sim R/\sin\psi=7\times 10^{15}$ cm for
a jet semi--opening angle $\psi =0.1$).
For these data we therefore prefer an inverse Compton region within 
the Broad Line Region (``IN" in table 4)
choice.
For the 1999 December observation we do not have tight constraints 
from variability,
and one should consider both regions within and outside the BLR. 

In any case, for both the June and December 1999 SEDs, we show the 
calculated spectra for both options, to illustrate the difference
in the predicted high energy spectra.

In Table 4 we list the relevant input parameters of the model.
As can be seen the different SEDs can be explained by using the same intrinsic 
luminosity, apart from the 1997 flare case (which has 5 times more power than the
other SEDs).
What largely differs among the different cases is the value of the magnetic field
and the value of $\gamma_{min}$, which controls the location of both the synchrotron
and the Compton peaks and their relative amplitude.

\section{Conclusions}
 
{\it Beppo}SAX observations of BL \,Lac in 1999 reveal the extremely 
complex behaviour of this source, as confirmed by  the comparison 
with previous multiwavelength observations.
During the June observation, soft X-ray light curves were 
characterized by the fastest variability episode ever recordered for BL \,Lac:
 the flux below 4 keV doubled in 20 minutes
while remaining constant at higher energies. 
Such an amazing event allows us to put severe constraints on the dimension
 of the X-ray emission region, on the magnetic field and on
 the emitting particle energies.
The frequency dependence is easily explained when performing
the spectral analysis, that highlights a spectral  break attributed to the
transition from the more variable synchrotron to a very hard 
inverse Compton spectrum: synchrotron X-ray emitting electrons are more
 energetic than Compton ones, so they cool faster. 
During the second 1999 run, {\it Beppo}SAX detected a softer Compton 
component along its whole spectral range, which accounts for the constance
of the light curves.
The comparison of 1995 and 1997 BL \,Lac Spectral Energy Distributions,
 extending to $\gamma$-ray energies,  
suggests the implication of different inverse Compton emission models. 
Furthermore, the constance of external radiation, 
as inferred from the constance 
of emission lines, implies that the emitting shell should be
 differently located  with respect to the Broad Line Region,
in order to explain the different observed high energy spectra. 
Unfortunately our observations lack simultaneous $\gamma$-ray data,
 which are essential 
to discriminate between various high energy emission models. 
Therefore, we have calculated the Spectral Energy Distribution resulting
 from different possible locations
of the emitting shell, adopting  an internal shock model. 
Taking into consideration
 the hints  given by variability and X-ray spectral shape,
the most viable scenario to explain our multiwavelength
 BL \,Lac observation of  1999  June implies that the emitting  shell
is internal to the Broad Line Region,
 thus producing $\gamma$-ray emission via Compton scattering 
with external photons.  
The 1999 December data are less revealing and 
we cannot discriminate between the two possible locations.

\acknowledgements{
This research was financially supported by the Italian Space 
Agency and MURST. We thank the {\it Beppo}SAX Science Data
Center (SDC) for their support in the data analysis.
This research made use of the NASA/IPAC Etragalactic Database (NED)
which is operated by the Jet Propulsion Laboratory, Caltech, under
contract with the National Aeronautics and Space Administration.}

\end{document}